\documentclass[runningheads]{llncs}

\usepackage[table,dvipsnames]{xcolor}

\usepackage{graphicx}

\usepackage{booktabs}
\usepackage{multirow}
\usepackage{colortbl}

\usepackage{pifont}

\usepackage{realboxes}     
\usepackage{algorithm}
\usepackage{algorithmic}
\usepackage{amsmath,amsfonts}
\usepackage{multicol}
\usepackage{subfig} 
\begin{document}
\title{
Attackers Strike Back? Not Anymore — An Ensemble of RL Defenders Awakens for APT Detection.}
\titlerunning{An Active Adversarial Multi-Agent RL Framework for APT Detection.}
%
\author{%
Sidahmed Benabderrahmane \and
Talal Rawhan
}%
\authorrunning{S. Benabderrahmane et al.}
\institute{
New York University, NYUAD, Division of Science.\\
\email{sidahmed.benabderrahmane@gmail.com}
}

\maketitle              

\begin{abstract}
Advanced Persistent Threats (APTs) represent a growing menace to modern digital infrastructure. Unlike traditional cyberattacks, APTs are stealthy, adaptive, and long-lasting, often bypassing signature-based detection systems. This paper introduces a novel framework for APT detection that unites deep learning, reinforcement learning (RL), and active learning into a cohesive, adaptive defense system. Our system combines autoencoders for latent behavioral encoding with a multi-agent ensemble of RL-based defenders, each trained to distinguish between benign and malicious process behaviors.

We identify a critical challenge in existing detection systems: their static nature and inability to adapt to evolving attack strategies. To this end, our architecture includes multiple RL agents (Q-Learning, PPO, DQN, adversarial defenders), each analyzing latent vectors generated by an autoencoder. When any agent is uncertain about its decision, the system triggers an active learning loop to simulate expert feedback, thus refining decision boundaries. An ensemble voting mechanism, weighted by each agent’s performance, ensures robust final predictions.

 \end{abstract}
 
\section{Introduction}
 
\subsection{Overview:}

Cybersecurity is undergoing a fundamental transformation as Advanced Persistent Threats (APTs) emerge as some of the most insidious forms of cyber attacks \cite{Stuxnet11,marczak2018hide,saad2020attribution}. Unlike conventional threats, APTs are characterized by their stealth, persistence, and long-term objectives. They typically unfold over multiple phases—initial reconnaissance, exploitation, lateral movement, data exfiltration, and persistence—making them difficult to detect using traditional security systems \cite{ghafir2014advanced,shackelford2016protecting}.

The increased sophistication of APTs presents a profound challenge to defenders \cite{han2018tapp,jenkinson2017applying}. Signature-based Intrusion Detection Systems (IDS) fail to detect novel attack vectors, while classical anomaly detection systems struggle with high false positive rates and poor adaptability \cite{SALIM2023e17156,tankard2011advanced}. Moreover, APTs often mimic legitimate user behavior, leveraging living-off-the-land binaries (LOLBins) or hijacking system processes, further complicating detection.

To address these issues, there is growing interest in intelligent, learning-based defense mechanisms. Recent progress in machine learning has shown the potential of deep learning and reinforcement learning (RL) to model complex system behavior and adapt dynamically to evolving threats \cite{Nguyen23}. In particular, deep autoencoders have proven effective for reducing high-dimensional event data into compact latent representations that preserve essential patterns for anomaly detection \cite{article_1237742}.

Reinforcement learning introduces an agent-based paradigm wherein models learn optimal behaviors through interaction with an environment \cite{ADAWADKAR2022105116}. In the context of APT detection, RL agents can learn to classify process traces as benign or malicious by receiving feedback in the form of rewards. This approach is inherently adaptive, enabling defenders to adjust strategies based on new observations.

However, using a single RL agent is often insufficient. APTs exhibit diverse behaviors that challenge any single learning paradigm. Therefore, we propose an ensemble of heterogeneous RL agents, including Q-learning, PPO, DQN, and adversarially trained models, each contributing its strengths to the detection process. This ensemble not only increases robustness but also enables better generalization across varying attack scenarios.

To further refine the decision-making process, we integrate an active learning loop. When an RL agent encounters a sample about which it is uncertain—typically when outputs are near 0.5—it triggers a simulated query for expert labeling. This simulates real-world scenarios where human analysts are involved in labeling uncertain samples. By incorporating this feedback, agents are retrained iteratively to improve accuracy and reduce uncertainty.
 
\subsection{Problem Statement and Objective}
 
Advanced Persistent Threat (APT) detection remains a critical and unresolved challenge in cybersecurity. This difficulty stems from several factors: the rarity and subtlety of APT events in real-world datasets, the evolving strategies of attackers that induce concept drift over time, and the significant cost associated with false positives in sensitive operational environments.

To address these challenges, this paper introduces a multi-agent ensemble framework that integrates deep autoencoding, reinforcement learning, and active feedback loops. The framework is designed to extract latent patterns of system behavior through the use of autoencoders, while reinforcement learning agents are trained to identify APT activities using reward-driven learning paradigms. Additionally, an active learning mechanism is employed to concentrate model adaptation on uncertain and high-risk samples, effectively simulating expert feedback and accelerating learning. 

The final decision process aggregates the outputs of multiple agents through a weighted ensemble strategy, ensuring robustness and adaptability. By unifying unsupervised feature extraction, multi-agent reinforcement learning, and uncertainty-aware querying, this work presents a resilient and continuously evolving defense system capable of countering sophisticated APTs.

\section{Related Work}
 
Reinforcement Learning (RL) has emerged as a powerful paradigm in machine learning, characterized by an agent that learns to take actions in an environment to maximize a cumulative reward \cite{milani2024explainable,tang2025deep}. The RL framework is formally defined by a Markov Decision Process (MDP), represented as a tuple \((\mathcal{S}, \mathcal{A}, P, R, \gamma)\), where: \(\mathcal{S}\) is the set of states, \(\mathcal{A}\) is the set of actions, \(P(s'|s, a)\) defines the transition probabilities, \(R(s, a)\) is the reward function,  \(\gamma \in [0, 1]\) is the discount factor.

The goal is to learn a policy \(\pi(a|s)\) that maximizes the expected return:
\[ J(\pi) = \mathbb{E}_{\pi}\left[ \sum_{t=0}^{\infty} \gamma^t R(s_t, a_t) \right] \]

\subsection{Value-Based Methods: Q-Learning and DQN}
 
Q-Learning is a model-free algorithm that learns the optimal action-value function \(Q^*(s, a)\) using the Bellman equation:
\[ Q(s_t, a_t) \leftarrow Q(s_t, a_t) + \alpha \left[ r_t + \gamma \max_{a'} Q(s_{t+1}, a') - Q(s_t, a_t) \right] \]

Deep Q-Networks (DQN) extend Q-learning by using a neural network to approximate \(Q(s, a; \theta)\), where \(\theta\) are learnable parameters. The loss function is: $\mathcal{L}(\theta) = \left( r_t + \gamma \max_{a'} Q(s_{t+1}, a'; \theta^-) - Q(s_t, a_t; \theta) \right)^2$, where \(\theta^-\) denotes a target network.
 
\subsection{Policy-Based Methods: REINFORCE and PPO}

In policy-based methods, we directly learn a parameterized policy \(\pi_\theta(a|s)\). The REINFORCE algorithm updates the policy via gradient ascent on the expected return:
$ \nabla_\theta J(\theta) = \mathbb{E}_{\pi_\theta} \left[ \nabla_\theta \log \pi_\theta(a|s) \cdot R_t \right] $

Proximal Policy Optimization (PPO) improves on REINFORCE by using a clipped surrogate objective:
$ L(\theta) = \mathbb{E} \left[ \min \left( r(\theta) \hat{A}, \text{clip}(r(\theta), 1 - \epsilon, 1 + \epsilon) \hat{A} \right) \right] $
where \(r(\theta) = \frac{\pi_\theta(a|s)}{\pi_{\theta_{old}}(a|s)}\) is the probability ratio and \(\hat{A}\) is the advantage estimate.

\subsection{Adversarial RL}
 
Adversarial reinforcement learning introduces a perturbation-generating agent that seeks to fool the defender agent. This setting is often modeled as a minimax game: $ \min_{\pi_D} \max_{\pi_A} \mathbb{E}[ R(s, a_D, a_A) ]$, 
where \(\pi_D\) is the defender’s policy and \(\pi_A\) is the attacker’s.
 
\subsection{Multi-Agent and Ensemble RL}
 
Multi-agent reinforcement learning (MARL) extends single-agent RL to environments with multiple agents, which may be cooperative, competitive, or mixed. The joint policy \(\pi = (\pi_1, \dots, \pi_n)\) is learned to optimize a shared or individual reward signal.

Ensemble reinforcement learning combines predictions from multiple agents. In our work, we use weighted voting based on AUC performance:
\[ P_{\text{ensemble}}(APT) = \sum_{i=1}^{N} w_i P_i(\text{APT}) \quad \text{with} \quad w_i \propto \text{AUC}_i \]

Few studies in cybersecurity combine these elements. Our contribution lies in synthesizing RL fundamentals with active learning feedback and ensemble predictions for robust, adaptive APT detection.
 
\section{Proposed Method}
 
Our proposed method introduces a multi-agent reinforcement learning (RL) framework combined with active learning and autoencoder-based latent encoding for robust and adaptive APT detection. The primary components and contributions of our architecture are as follows:
 
\begin{itemize}
    \item We use a deep autoencoder to learn compact, latent representations of system behavior traces, allowing us to isolate meaningful behavioral anomalies.
    \item We train a set of heterogeneous RL agents (Q-Learning, PPO, DQN, Adversarial Agents), each making decisions independently based on latent features and AE reconstruction error.
    \item We implement an active learning loop that identifies uncertain predictions using softmax-based confidence margins, queries the Oracle (e.g. ground truth), and updates the RL agents accordingly.
    \item We introduce a weighted ensemble strategy that aggregates the decisions of all agents using their AUC scores as confidence weights.
    \item We propose a feedback refinement loop for the autoencoder by incorporating false-positive benign feedback, simulating analyst verification.
\end{itemize}

\subsection{Autoencoder-based Latent Representation}
 
To efficiently detect Advanced Persistent Threats (APTs), which are often buried within high-dimensional system activity logs, we utilize an attention-based autoencoder AAE to derive compact and informative latent representations of process behavior. Let each input sample be a binary vector \( x \in \{0,1\}^d \), representing the presence or absence of specific system-level events or features. The autoencoder is a neural network composed of two parts: an encoder \( f_{enc} \) and a decoder \( f_{dec} \).

\begin{equation}
    z = f_{enc}(x; \theta_{enc}) \in \mathbb{R}^k, \quad \text{with } k \ll d
\end{equation}

\begin{equation}
    \hat{x} = f_{dec}(z; \theta_{dec}) \in \mathbb{R}^d
\end{equation}

Here, \( z \) denotes the latent vector capturing compressed semantic information from \( x \), and \( \hat{x} \) is the reconstruction.

The reconstruction error, defined as the mean squared error (MSE) between the input and its reconstruction, is given by: $  \mathcal{L}_{AE}(x, \hat{x}) = \frac{1}{d} \sum_{i=1}^{d} (x_i - \hat{x}_i)^2$.  The autoencoder is trained to minimize this loss:
$\min_{\theta_{enc}, \theta_{dec}} \ \mathbb{E}_{x \sim \mathcal{D}}[ \mathcal{L}_{AE}(x, f_{dec}(f_{enc}(x))) ]$. The reconstruction error \( \mathcal{L}_{AE} \) serves two purposes: (i) As an anomaly score: samples with high reconstruction error are likely anomalous. (ii) As a reward signal: to guide reinforcement learning agents (discussed in later sections). We denote the latent representation for the entire dataset as: $Z = \{ z_i = f_{enc}(x_i) \}_{i=1}^{N}$, \text{for } N \text{ samples}. These latent embeddings are used as the input state representations for the reinforcement learning agents, enabling them to operate in a reduced-dimensional, semantically meaningful space.
 
\subsection{Multi-Agent RL Architecture}
 
To capture the sequential and interactive aspects of APT detection, we employ multiple reinforcement learning (RL) agents that act as autonomous defenders. Each agent receives a state derived from the AAE autoencoder  latent space and outputs a prediction indicating whether the behavior is benign or malicious. We use multiple types of RL agents: Q-Learning, DQN, PPO, Adversarial Agent, Multiple Agents, and Active Adversarial Agents.

Let the environment be defined as a tuple \( (\mathcal{S}, \mathcal{A}, R, \mathcal{T}, \gamma) \), where: \( \mathcal{S} \): Set of states (AAE autoencoder latent vectors \( z \)). \( \mathcal{A} = \{0,1\} \): Actions (0 = benign, 1 = APT). \( R(s, a) \): Reward function based on prediction correctness and AAE error.  \( \mathcal{T}(s, a, s') \): State transition dynamics (stateless in this context).  \( \gamma \in [0,1) \): Discount factor.

Each agent is trained on the latent space output of the autoencoder. The agents maintain or approximate a Q-function \( Q(s, a) \) to guide decision-making:
\begin{equation}
    Q^*(s, a) = \mathbb{E} \left[ \sum_{t=0}^{\infty} \gamma^t R(s_t, a_t) \mid s_0 = s, a_0 = a \right]
\end{equation}

\textbf{Q-Learning Agent:}
Uses a tabular or neural approximation of \( Q(s, a) \) updated via:
\begin{equation}
    Q(s_t, a_t) \leftarrow Q(s_t, a_t) + \alpha \left[ r_t + \gamma \max_a Q(s_{t+1}, a) - Q(s_t, a_t) \right]
\end{equation}

\textbf{DQN Agent:}
Approximates the Q-function with a deep network:
\begin{equation}
    \mathcal{L}_{\text{DQN}} = \mathbb{E}_{(s,a,r,s')} \left[ (r + \gamma \max_{a'} Q(s', a'; \theta^-) - Q(s, a; \theta))^2 \right]
\end{equation}
where \( \theta \) and \( \theta^- \) are the online and target network parameters.

\textbf{PPO Agent:}
Optimizes a stochastic policy \( \pi_\theta(a|s) \) using a clipped surrogate objective:
\begin{equation}
    L^{\text{CLIP}}(\theta) = \mathbb{E}_t \left[ \min\left( r_t(\theta) \hat{A}_t, \text{clip}(r_t(\theta), 1 - \epsilon, 1 + \epsilon) \hat{A}_t \right) \right]
\end{equation}
where \( r_t(\theta) = \frac{\pi_\theta(a_t|s_t)}{\pi_{\theta_{\text{old}}}(a_t|s_t)} \) and \( \hat{A}_t \) is the estimated advantage.

\textbf{Adversarial Agent:}
Trained under input perturbations \( \delta \) to ensure robustness:
\begin{equation}
    \max_{\delta \in \Delta} \mathcal{L}(Q(s + \delta, a), y), \quad \text{subject to } \|\delta\| \leq \epsilon
\end{equation}


\textbf{Multi-Agent RL Architecture:}
To enhance detection capability and resilience, we adopt a multi-agent reinforcement learning (MARL) architecture wherein multiple heterogeneous RL agents operate in parallel, share training signals, and collaboratively improve their anomaly detection performance.

Let \( \mathcal{A}_{RL} = \{ A_1, A_2, \dots, A_n \} \) denote the set of RL agents. Each agent \( A_i \) receives a state \( s_i = [z_i, \epsilon_i] \), where \( z_i \in \mathbb{R}^k \) is the latent vector from the autoencoder, and \( \epsilon_i \in \mathbb{R} \) is the corresponding reconstruction error. Each agent aims to learn a policy \( \pi_i(a|s) \) or Q-function \( Q_i(s, a) \) to maximize its expected reward:
\begin{equation}
    \mathbb{E}_{s \sim \mathcal{S}, a \sim \pi_i} \left[ \sum_{t=0}^{\infty} \gamma^t R(s_t, a_t) \right]
\end{equation}

We define a unified reward function across agents:
\begin{equation}
    R(s, a) =
    \begin{cases}
    +1 + \beta \cdot \epsilon & \text{if correct prediction} \\
    -1 + \beta \cdot \epsilon & \text{if benign misclassified} \\
    -2 + \beta \cdot \epsilon & \text{if APT missed}
    \end{cases}
\end{equation}

Here, \( \beta \in \mathbb{R}^{+} \) is a tunable coefficient that modulates the influence of the reconstruction error \( \epsilon \) on the reward signal. This parameter plays a crucial role in shaping the behavior of the RL agents. It allows the reward to reflect not only the correctness of the prediction but also the degree of anomaly suggested by the autoencoder. A higher \( \beta \) increases the agent's sensitivity to uncertain or anomalous states, thereby directing more learning focus toward high-risk areas. This mechanism encourages agents to assign greater importance to difficult cases that exhibit unusual behavior, effectively balancing classification performance with anomaly intensity.

This architecture provides redundancy, robustness, and the ability to learn complementary patterns, ultimately enhancing the detection of rare and stealthy APT behaviors.
\vspace{-0.8 em}
\subsection{Active Learning with Reward-Shaped Retraining}
\vspace{-0.8 em}
To improve sample efficiency, focus learning on ambiguous cases, and address the scarcity of labeled anomalous samples, we incorporate a targeted active learning loop into the MARL framework. At each inference step, an agent assesses its confidence level on the current input. If the output is deemed uncertain (e.g., softmax near 0.5 or Q-values close), the corresponding state is flagged and sent for simulated oracle labeling.

Uncertainty is computed using the margin criterion:
\begin{equation}
    \text{uncertainty}(s) = |\pi(a=1|s) - 0.5| < \delta \quad \text{or} \quad |Q(s,1) - Q(s,0)| < \delta
\end{equation}

Queried samples \( (s, y) \) are stored in an active replay buffer and used to fine-tune the corresponding agent. The training loss incorporates a reward modification based on the AE error:
$ 
    R'(s, a) = R(s, a) + \lambda \cdot \mathcal{L}_{AE}(x, \hat{x})
$. This encourages agents to prioritize high-error regions, effectively guiding learning toward suspicious behavior. The overall active learning routine is summarized in Algorithm~\ref{alg:active-loop}.

\begin{algorithm}[H]
\caption{Active Learning with Reward-Shaped Retraining}
\begin{algorithmic}[1]
\FOR{each agent $A_i$}
    \FOR{each test sample $s$}
        \STATE Predict $Q(s, a)$ or $\pi(a|s)$
        \IF{uncertain($s$)}
            \STATE Simulate oracle to obtain $y$
            \STATE Add $(s, y)$ to active buffer
        \ENDIF
    \ENDFOR
    \STATE Retrain $A_i$ using buffer with reward $R'(s, a)$
\ENDFOR
\end{algorithmic}
\label{alg:active-loop}
\end{algorithm}

This loop is executed at each feedback iteration. The impact is evaluated by computing model AUC before and after updates. This feedback-driven process helps refine decision boundaries in low-confidence areas, increasing true positive rates while reducing labeling cost.

\subsection{Ensemble Decision Module}

To consolidate the diverse predictions of the multi-agent RL system, we employ an ensemble strategy that aggregates the outputs of individual agents into a single final decision. This enhances robustness and allows leveraging the strengths of different models.

Let \( \{ A_1, A_2, ..., A_n \} \) be the set of agents, each producing a score or class prediction \( y_i \in \{0, 1\} \). We explore following ensemble strategy:

\textbf{1. Majority Voting:}
\begin{equation}
    y_{final} = \arg\max_{y \in \{0, 1\}} \sum_{i=1}^{n} \mathbb{1}(y_i = y)
\end{equation}

This decision fusion step capitalizes on diversity among agents and has been shown empirically to improve detection of stealthy APTs across datasets. 

\section{Results}
\subsection{Dataset Description}
We use real-world APT datasets with binary process behavior vectors. Each vector represents the presence/absence of key system actions. APTs are rare and labeled via known attack stages (recon, exploit, persist, etc.). These cyber security data sources used in this paper come from the Defense Advanced Research Projects Agency (DARPA)’s \verb|Transparent| \verb|Computing TC|\footnote{https://gitlab.com/adaptdata} program~\cite{darpa},\cite{berrada_2019}, \cite{BerradaCBMMTW20},\cite{Benabderrahmane21},\cite{DBLP:journals/corr/abs-2006-07916},\cite{DBLP:journals/fgcs/BenabderrahmaneHVCR24}. The aim of this program is to provide transparent provenance data of system activities and component interactions across different operating systems (OS) and spanning all layers of software abstractions. Specifically, the datasets include system-level data, background activities, and system operations recorded while APT-style attacks are being carried out on the underlying systems. Preserving the provenance of all system elements allows for tracking the interactions and dependencies among components. Such an interdependent view of system operations is helpful for detecting activities that are individually legitimate or benign but collectively might indicate abnormal behavior or malicious intent.

Here we specifically employ DARPA’s data that has undergone processing conducted by the \verb|ADAPT| (Automatic Detection of Advanced Persistent Threats) project’s ingester~\cite{berrada_2019}, \cite{BerradaCBMMTW20}, \cite{Benabderrahmane21}, \cite{DBLP:journals/corr/abs-2006-07916}, \cite{DBLP:journals/fgcs/BenabderrahmaneHVCR24}. The records come from four different source OS, namely Android (called in the TC program Clearscope), Linux (called Trace), BSD (called Cadets), and Windows (called Fivedirections or 5dir). For each system, the data comes from two separate attack scenarios: scenario 1 (also called Pandex) and scenario 2 (called Bovia), respectively. The processing includes ingesting provenance graph data into a graph database as well as additional data integration and deduplication steps. The final data includes a number of Boolean-valued datasets (data aspects)
, with each representing an aspect of the behavior of system processes as illustrated in Table \ref{dataexample}. Each row in such a data aspect 
is a data point representing a single process run on the respective OS. It is expressed as a Boolean vector whereby a value of 1 in a vector cell indicates the corresponding attribute applies to that process. 


\begin{table}[!th]
\small
\begin{tabular}{|l|c|c|c|c|c|c|c|}
\hline
                                           & {\rotatebox{70}{/usr/sbin/avahi-autoipd}} & {\rotatebox{70}{216.73.87.152}}   & {\rotatebox{70}{EVENT\_OPEN} }    & {\rotatebox{70}{EVENT\_EXECUTE}}  & {\rotatebox{70}{EVENT\_CONNECT} } & {\rotatebox{70}{EVENT\_SENDMSG} } & \multicolumn{1}{l|}{...} \\ \hline
{ee27fff2-a0fd-1f516db3d35f} & 1                                & 1                        & 1                        & 0                        & 1                        & 0                        & ...                      \\ \hline
{b2e7e930-8f25-4242a52c5d72} & 0                                & 1                        & 0                        & 1                        & 1                        & 1                        & ...                      \\ \hline
{07141a2a-832e-8a71ca767319} & 0                                & 0                        & 1                        & 1                        & 1                        & 1                        & ...                      \\ \hline
{b4be70a9-98ac-81b0042dbecb} & 1                                & 0                        & 1                        & 1                        & 0                        & 0                        & ...                      \\ \hline
{2bc3b5c6-9110-076710a13038} & 0                                & 0                        & 0                        & 0                        & 0                        & 1                        & ...                      \\ \hline
{ad7716e0-8d59-5d45d1742211} & 1                                & 1                        & 0                        & 1                        & 0                        & 1                        & ...                      \\ \hline
...                                           & \multicolumn{1}{l|}{...}         & \multicolumn{1}{l|}{...} & \multicolumn{1}{l|}{...} & \multicolumn{1}{l|}{...} & \multicolumn{1}{l|}{...} & \multicolumn{1}{l|}{...} & \multicolumn{1}{l|}{...} \\ \hline
\end{tabular}
\caption{Example of a boolean-valued dataset (data aspect) from the DARPA TC program. In each row, the boolean vector represents the list of features of the corresponding process. }
 \label{dataexample}
 \vspace{-2.5 em}
\end{table}
For instance, in Table \ref{dataexample}, the process with id\\ \verb|ee27fff2-a0fd-1f516db3d35f| has the following sequence of events:\\
\verb|</usr/sbin/avahi-autoipd|, \verb|216.73.87.152|, \verb|EVENT_OPEN, EVENT_CONNECT, ...>|. \\
Specifically, the relevant 
datasets are interpreted as follows:

\begin{itemize}
    \item \verb|ProcessEvent| (PE): Its attributes are event types performed by the processes. A value of 1 in \verb|process[i]| means the process has performed at least one event of type $i$.
    \item \verb|ProcessExec| (PX): The attributes are executable names that are used to start the processes.
    \item \verb|ProcessParent| (PP): Its attributes are executable names that are used to start the parents of the processes.
    \item \verb|ProcessNetflow| (PN): The attributes here represent IP addresses and port names that have been accessed by the processes.
    \item \verb|ProcessAll| (PA): This dataset is described by the disjoint union of all attribute sets from the previous datasets.
\end{itemize}
Overall, with two attack scenarios, four OS (BSD, Windows, Linux, Android) and five aspects (PE, PX, PP, PN, PA), a total of forty individual datasets are composed, as illustrated in Figure 1. 
\vspace{-1 em}
\begin{figure}
    \centering
    \includegraphics[width=0.4\linewidth]{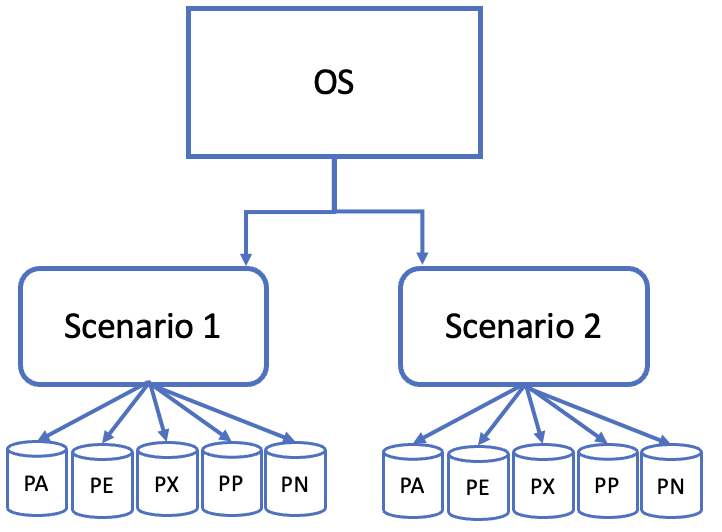}
    \caption{Organization of the DARPA's TC datasets. Each OS undergoes two attack scenarios, each of which contains five data aspects ets. With four OS (BSD, Windows, Linux, Android), two attack scenarios, and five aspects (PE, PX, PP, PN, PA), a total of forty individual datasets are composed. Each forensic configuration (OS$\times$attack scenario$\times$data aspect) represents a single dataset \cite{DBLP:journals/fgcs/BenabderrahmaneHVCR24}.}
    \label{fig:DARPATC}
    \vspace{-1.9 em}
\end{figure}
They are described in Table~\ref{datatable} whereby the last column provides the number of attacks in each dataset. The substantially imbalanced nature of the datasets is clearly seen here. For each forensic configuration (OS$\times$attack scenario$\times$data aspect) we have the number of processes (instances) and the corresponding events (features). For instance, $Windows\_E1\_PE$ is the dataset represented by $PE$ aspect belonging to Windows OS, produced during the first attack scenario E1 (Pandex). It contains 17569 instances and 22 features with a total of 8 APTs anomalies (0.04\%). 
\vspace{-0.8 em}

\begin{table*}
\centering
\resizebox{0.99\textwidth}{!}{
\begin{tabular}{|l|l||l|l|l|l|l|l|l|l|}
\hline & Scenario & Size& $PE$   & $PX$  & $PP$  & $PN$     & $PA$  & $nb\_attacks$    & $\%\frac{nb\_attacks}{nb\_processes}$     \\ \hline \hline
BSD    & 1 &288 MB &76903 / 29  & 76698 / 107  & 76455 / 24  & 31 / 136  & 76903 / 296 & 13&0.02\\  
    & 2 &1.27 GB &224624 / 31  &224246 / 135  & 223780 / 37  & 42888 / 62 &  224624 / 265      & 11&0.004\\ \hline
Windows & 1 &743 MB & 17569 / 22    &  17552 / 215  &   14007 / 77        &   92 / 13963      & 17569 / 14431& 8&0.04\\  
   & 2 &9.53 GB& 11151 / 30    &  11077 / 388  & 10922 / 84  & 329 / 125      &  11151 / 606    &8&0.07\\ \hline
Linux  & 1 &2858 MB &247160 / 24 & 186726 / 154 & 173211 / 40 & 3125 / 81 & 247160 / 299  &25&0.01\\
    & 2 &25.9 GB &282087 / 25 & 271088 / 140 & 263730 / 45 &6589 / 6225 &  282104 / 6435      &46&0.01\\ \hline
Android& 1 &2688 MB&102 / 21     &102 / 42&0 / 0&8 / 17& 102 / 80&9&8.8\\
&2 &10.9 GB&12106 / 27     &12106 / 44&0 / 0&4550 / 213&12106 / 295 &13&0.10\\ \hline
\end{tabular}
}
\caption{Summary of the first source of 40 benchmark datasets belonging to DARPA's TC program for APT detection. A dataset entry (columns 4 to 8) is described by a number of rows (processes) / number of columns (attributes). For instance, with ProcessAll (PA) obtained from the second scenario using Linux, the dataset has 282104 rows and 6435 attributes with 46 APT attacks (0.01\%). }
 \label{datatable}
 \vspace{-2 em}
\end{table*}
\subsection{Results and Discussion}
\vspace{-0.8 em}
In this section, we present and analyze the performance of our proposed active adversarial multi-agent reinforcement learning (RL) framework for Advanced Persistent Threat (APT) detection. We structure the discussion around the following key aspects:
\begin{itemize}
    \item \textit{Evaluation Setup:} We evaluate our framework using real-world datasets containing labeled traces of benign and APT activities. The models are tested on a separate validation set after each active learning iteration. Evaluation metrics include Area Under the Curve (AUC), Precision, Recall and F1-score to assess classification performance and ranking quality.
    \item \textit{Performance of Individual RL Agents:} We compare the effectiveness of Q-Learning, DQN, PPO, and an Adversarial Agent in detecting APT behavior. Each model is trained on latent representations derived from the autoencoder and retrained using active feedback. Their performance is tabulated over 50 iterations of active learning.
\item \textit{Active Learning Impact:} We analyze the influence of the active learning loop by tracking the AUC scores across iterations. Results show a consistent improvement in true positive rates and a reduction in false positives. Queried samples effectively refine the decision boundaries of the agents.
\item \textit{Ensemble Detection Strategy:} We evaluate the ensemble strategy with a majority voting and aggregation. 
\end{itemize}
\vspace{-0.8 em}
\subsubsection{Evaluated Models}
\vspace{-0.8 em}
In our study, we evaluate the effectiveness of eight core anomaly detection models—seven based on reinforcement learning (RL) and one based on unsupervised reconstruction. These models are integrated into a broader multi-agent and ensemble-based architecture aimed at detecting Advanced Persistent Threats (APTs). Each model processes system behavior in the latent space produced by the AAE autoencoder and returns predictions indicating whether the behavior is benign or malicious.

\begin{itemize}
    \item \textit{Q-Learning Agent (Q-RL):} A classical value-based agent that updates its Q-values through tabular or function approximation using the Bellman equation. It serves as a baseline reinforcement learner with discrete state-action feedback.
    
    \item \textit{Deep Q-Network (DQN):} An extension of Q-Learning that employs a deep neural network to approximate the Q-function. It uses experience replay and target networks to stabilize training in high-dimensional spaces.
    
    \item \textit{Proximal Policy Optimization (PPO):} A policy-gradient agent that optimizes a clipped surrogate objective to ensure stable updates. PPO is known for balancing exploration and exploitation efficiently.
        \item \textit{Rule-Based Autoencoder (AE-RL):} A baseline model that flags a process as malicious if its reconstruction error exceeds a predefined threshold. It serves as an unsupervised anomaly detector leveraging learned representations.
    \item \textit{Adversarial Agent (AMARL):} A Q-learning-based model trained on both clean and perturbed inputs to ensure robustness against evasion attacks. The adversarial training process encourages the agent to remain stable under adversarial manipulations.

    \item  \textit{Active Learning Adversarial Agent (AAMARL):}
To assess the impact of uncertainty-driven active learning, we simulate an iterative feedback loop over 50 rounds. During each iteration, the model identifies uncertain samples based on the softmax margin of its predictions. These samples are considered candidates for querying an oracle, which returns the ground truth labels. The newly labeled instances are then incorporated into the training set, enabling the model to refine its decision boundary incrementally.

After each round of feedback and retraining, we evaluate the model on the full test set, computing the AUC and F1-score. This process yields a trajectory of performance metrics over the active learning timeline. The final reported AUC and F1 for the ``Active Multi-Agent'' configuration in Table~3 represent the average performance across all 50 iterations, thereby reflecting the cumulative benefit of integrating human feedback into the learning pipeline.

\item \textit{Meta Active Adversarial Multi-Agent Evaluation (EAAMARL):}
In addition to individual agent retraining, we deploy active learning in the context of our proposed ensemble model, which combines predictions from Q-Learning, PPO, DQN, and adversarially trained agents. Over 50 active learning iterations, each agent selectively queries uncertain samples based on its softmax output margin. Ground truth labels for these samples are simulated as oracle feedback and integrated into each agent's experience replay buffer.

As agents improve individually through this iterative process, their collective predictions—aggregated using AUC-weighted voting—are updated accordingly. After each feedback round, the ensemble’s decision is evaluated on the full test set using AUC and F1-score. The average values over the 50 iterations are reported as the final performance for the ``EAAMARL'' row in Table~\ref{tab:simulated-results}, highlighting the gains achieved by coordinated uncertainty-aware learning across multiple adversarially hardened agents.

\end{itemize}

Each of these models is integrated into an ensemble for improved robustness, and further enhanced using an active learning loop to address uncertainty in prediction and guide retraining dynamically.

Each model processes a latent vector obtained from the autoencoder and aims to classify whether the input behavior corresponds to a benign process or an APT. The agents were evaluated across 40 different datasets constructed by combining the following:
\vspace{-0.8 em}
\begin{itemize}
    \item \textit{Operating Systems (4):} BSD, Windows, Linux, Android
    \item \textit{Attack Scenarios (2):} Scenario 1, Scenario 2
    \item \textit{Datasets (5):} PA, PE, PX, PN, PP
\end{itemize}
\vspace{-0.8 em}

This results in a total of \textit{40 datasets}  allowing a comprehensive assessment of model performance across heterogeneous environments.
\begin{table}

\centering
\caption{AUC and F1 Score Comparison Across Reinforcement Learning-Based APT Detection Models. Each entry reports the AUC and F1 score (\textit{AUC / F1}) achieved on the corresponding dataset. The evaluated models include: Q-Learning (Q-RL), Proximal Policy Optimization (PPO), Deep Q-Network (DQN), Autoencoder-guided Reinforcement Learning (AE-RL), Multi-Agent Reinforcement Learning (MARL), Adversarial Multi-Agent Reinforcement Learning (AMARL), Active Adversarial Multi-Agent Reinforcement Learning (AAMARL), and Ensemble Active Adversarial Multi-Agent Reinforcement Learning (EAAMARL). The best-performing models are highlighted by consistently higher AUC and F1 scores across diverse datasets and attack scenarios.}

\label{tab:simulated-results}
\Rotatebox{90}{
\scriptsize

\begin{tabular}{llc||c|c|c|c|c|c|c|c}
\hline
\multicolumn{3}{l|}{\textbf{Database (OS $\times$ Scenario $\times$ Dataset)}} & \textbf{Q-RL} & \textbf{PPO} & \textbf{DQN} & \textbf{AE-RL} & \textbf{MARL} & \textbf{AMARL} & \textbf{AAMARL}      & \textbf{EAAMARL}     \\ \hline
\multirow{10}{*}{BSD}              & \multirow{5}{*}{E1}          & PA         & 0.57 / 0.17   & 0.68 / 0.21  & 0.65 / 0.12  & 0.70/ 0.19     & 0.74 / 0.26   & 0.78 / 0.33    & \textbf{0.85 / 0.42} & \textbf{0.91 / 0.50} \\
                                   &                              & PE         & 0.56 / 0.15   & 0.63 / 0.19  & 0.63 / 0.12  & 0.69 / 0.15    & 0.71 / 0.20   & 0.74 / 0.30    & 0.83 / 0.39          & \textbf{0.92 / 0.50} \\
                                   &                              & PX         & 0.51 / 0.01   & 0.58 / 0.20  & 0.58 / 0.10  & 0.65 / 0.14    & 0.69 / 0.14   & 0.73 / 0.28    & 0.82 / 0.38          & 0.85 / 0.43          \\
                                   &                              & PN         & 0.51 / 0.10   & 0.56 / 0.16  & 0.55 / 0.11  & 0.66 / 0.10    & 0.71 / 0.16   & 0.68 / 0.22    & 0.84 / 0.33          & 0.82 / 0.38          \\
                                   &                              & PP         & 0.55 / 0.13   & 0.55 / 0.16  & 0.51 / 0.10  & 0.62 / 0.07    & 0.67 / 0.15   & 0.63 / 0.21    & 0.81 / 0.26          & 0.82 / 0.36          \\ \cline{2-11} 
                                   & \multirow{5}{*}{E2}          & PA         & 0.56 / 0.13   & 0.63 / 0.19  & 0.64 / 0.12  & 0.69 / 0.17    & 0.69 / 0.16   & 0.76 / 0.28    & 0.83 / 0.33          & 0.88 / 0.45          \\
                                   &                              & PE         & 0.55 / 0.08   & 0.60 / 0.19  & 0.60 / 0.11  & 0.65 / 0.09    & 0.72 / 0.15   & 0.74 / 0.22    & \textbf{0.84 / 0.31} & \textbf{0.89 / 0.46} \\
                                   &                              & PX         & 0.51 / 0.01   & 0.57 / 0.15  & 0.55 / 0.08  & 0.66 / 0.10    & 0.73 / 0.16   & 0.68 / 0.19    & 0.82 / 0.32          & 0.86 / 0.37          \\
                                   &                              & PN         & 0.50 / 0.02   & 0.53 / 0.10  & 0.54 / 0.07  & 0.63 / 0.08    & 0.68 / 0.13   & 0.68 / 0.21    & 0.82 / 0.28          & 0.83 / 0.40          \\
                                   &                              & PP         & 0.51 / 0.08   & 0.51 / 0.11  & 0.52 / 0.08  & 0.62 / 0.07    & 0.67 / 0.12   & 0.68 / 0.20    & 0.81 / 0.27          & 0.82 / 0.30          \\ \hline
\multirow{10}{*}{Windows}          & \multirow{5}{*}{E1}          & PA         & 0.54 / 0.19   & 0.55 / 0.12  & 0.58 / 0.12  & 0.67 / 0.11    & 0.72 / 0.17   & 0.69 / 0.29    & \textbf{0.87 / 0.35} & \textbf{0.91 / 0.41} \\
                                   &                              & PE         & 0.56 / 0.02   & 0.55 / 0.12  & 0.56 / 0.12  & 0.64 / 0.08    & 0.69 / 0.14   & 0.69 / 0.29    & 0.83 / 0.30          & 0.88 / 0.36          \\
                                   &                              & PX         & 0.53 / 0.08   & 0.52 / 0.16  & 0.54 / 0.10  & 0.65 / 0.09    & 0.70 / 0.15   & 0.70 / 0.23    & 0.85 / 0.34          & 0.89 / 0.39          \\
                                   &                              & PN         & 0.57 / 0.19   & 0.56 / 0.15  & 0.52 / 0.09  & 0.63 / 0.08    & 0.68 / 0.13   & 0.69 / 0.19    & 0.82 / 0.30          & 0.85 / 0.35          \\
                                   &                              & PP         & 0.53 / 0.08   & 0.53 / 0.11  & 0.50 / 0.06  & 0.61 / 0.07    & 0.67 / 0.12   & 0.68 / 0.18    & 0.80 / 0.25          & 0.81 / 0.32          \\ \cline{2-11} 
                                   & \multirow{5}{*}{E2}          & PA         & 0.60/ 0.23    & 0.57 / 0.17  & 0.56 / 0.12  & 0.65 / 0.11    & 0.70 / 0.15   & 0.69 / 0.22    & \textbf{0.84 / 0.37} & \textbf{0.88 / 0.37} \\
                                   &                              & PE         & 0.57 / 0.12   & 0.51 / 0.12  & 0.53 / 0.10  & 0.64 / 0.08    & 0.69 / 0.14   & 0.69 / 0.19    & 0.83 / 0.33          & \textbf{0.89 / 0.39} \\
                                   &                              & PX         & 0.51 / 0.07   & 0.52 / 0.09  & 0.52 / 0.08  & 0.66 / 0.10    & 0.71 / 0.16   & 0.68 / 0.16    & 0.83 / 0.32          & 0.89 / 0.37          \\
                                   &                              & PN         & 0.52 / 0.08   & 0.52 / 0.09  & 0.52 / 0.08  & 0.63 / 0.08    & 0.68 / 0.13   & 0.63 / 0.15    & 0.82 / 0.27          & 0.84 / 0.33          \\
                                   &                              & PP         & 0.50 / 0.09   & 0.53 / 0.11  & 0.54 / 0.10  & 0.62 / 0.07    & 0.67 / 0.13   & 0.53 / 0.12    & 0.81 / 0.27          & 0.84 / 0.31          \\ \hline
\multirow{10}{*}{Linux}            & \multirow{5}{*}{E1}          & PA         & 0.55 / 0.16   & 0.56 / 0.15  & 0.69 / 0.13  & 0.66 / 0.19    & 0.70 / 0.15   & 0.66 / 0.22    & \textbf{0.85 / 0.39} & \textbf{0.89 / 0.36} \\
                                   &                              & PE         & 0.52 / 0.09   & 0.54 / 0.10  & 0.59 / 0.10  & 0.63 / 0.12    & 0.68 / 0.13   & 0.69 / 0.26    & 0.83 / 0.38          & 0.87 / 0.33          \\
                                   &                              & PX         & 0.53 / 0.03   & 0.53 / 0.06  & 0.55 / 0.10  & 0.65 / 0.09    & 0.70 / 0.15   & 0.70 / 0.22    & 0.84 / 0.31          & 0.88 / 0.35          \\
                                   &                              & PN         & 0.52 / 0.08   & 0.53 / 0.08  & 0.53 / 0.09  & 0.63 / 0.08    & 0.69 / 0.13   & 0.68 / 0.21    & 0.81 / 0.30          & 0.86 / 0.32          \\
                                   &                              & PP         & 0.50 / 0.06   & 0.51 / 0.05  & 0.51 / 0.06  & 0.62 / 0.07    & 0.67 / 0.16   & 0.63 / 0.21    & 0.80 / 0.29          & 0.84 / 0.34          \\ \cline{2-11} 
                                   & \multirow{5}{*}{E2}          & PA         & 0.56 / 0.10   & 0.57 / 0.08  & 0.58 / 0.09  & 0.66 / 0.13    & 0.70 / 0.16   & 0.72 / 0.21    & \textbf{0.83 / 0.39} & \textbf{0.87 / 0.33} \\
                                   &                              & PE         & 0.55 / 0.14   & 0.54/ 0.15   & 0.55 / 0.11  & 0.63 / 0.08    & 0.73 / 0.16   & 0.70 / 0.23    & 0.82 / 0.38          & 0.86 / 0.31          \\
                                   &                              & PX         & 0.52 / 0.08   & 0.52 / 0.08  & 0.52 / 0.12  & 0.65 / 0.09    & 0.69 / 0.15   & 0.69 / 0.22    & 0.80 / 0.30          & 0.81 / 0.35          \\
                                   &                              & PN         & 0.53 / 0.12   & 0.52 / 0.13  & 0.52 / 0.08  & 0.63 / 0.08    & 0.68 / 0.16   & 0.62 / 0.21    & 0.81 / 0.27          & 0.82 / 0.32          \\
                                   &                              & PP         & 0.51 / 0.08   & 0.50 / 0.09  & 0.51 / 0.09  & 0.62 / 0.07    & 0.67 / 0.12   & 0.60 / 0.19    & 0.80 / 0.25          & 0.81 / 0.30          \\ \hline
\multirow{8}{*}{Android}           & \multirow{4}{*}{E1}          & PA         & 0.55 / 0.09   & 0.55 / 0.10  & 0.57 / 0.11  & 0.65 / 0.09    & 0.72 / 0.22   & 0.63 / 0.22    & \textbf{0.85 / 0.36} & \textbf{0.89 / 0.36} \\
                                   &                              & PE         & 0.55 / 0.09   & 0.56 / 0.13  & 0.56 / 0.12  & 0.63 / 0.08    & 0.72 / 0.22   & 0.62 / 0.11    & 0.82 / 0.32          & 0.87 / 0.33          \\
                                   &                              & PX         & 0.51 / 0.10   & 0.52 / 0.10  & 0.53 / 0.08  & 0.70 / 0.19    & 0.71 / 0.15   & 0.58 / 0.12    & 0.84 / 0.33          & 0.84 / 0.35          \\
                                   &                              & PN         & 0.50 / 0.09   & 0.51 / 0.13  & 0.52 / 0.09  & 0.64 / 0.08    & 0.68 / 0.13   & 0.58 / 0.11    & 0.81 / 0.30          & 0.82 / 0.32          \\ \cline{2-11} 
                                   & \multirow{4}{*}{E2}          & PA         & 0.57 / 0.20   & 0.58 / 0.18  & 0.57 / 0.14  & 0.64 / 0.08    & 0.71 / 0.16   & 0.62 / 0.11    & 0.83 / 0.29          & 0.87 / 0.33          \\
                                   &                              & PE         & 0.52 / 0.13   & 0.53 / 0.14  & 0.55 / 0.12  & 0.63 / 0.12    & 0.70 / 0.19   & 0.59 / 0.16    & \textbf{0.85 / 0.28} & \textbf{0.88 / 0.41} \\
                                   &                              & PX         & 0.51 / 0.01   & 0.52 / 0.10  & 0.52 / 0.11  & 0.65 / 0.09    & 0.68 / 0.22   & 0.57 / 0.12    & 0.84 / 0.30          & 0.84 / 0.35          \\
                                   &                              & PN         & 0.54 / 0.14   & 0.55 / 0.15  & 0.51 / 0.09  & 0.63 / 0.08    & 0.66 / 0.18   & 0.58 / 0.16    & 0.81 / 0.27          & 0.82 / 0.32         
\end{tabular}
}
\end{table}

\vspace{-1 em}
\subsection{Discussion on Performance Accross Models}
\vspace{-0.8 em}
The results presented in Table~3 offer valuable insights into the comparative performance of reinforcement learning-based methods for Advanced Persistent Threat (APT) detection. 
\vspace{-0.8 em}
\paragraph{BSD:}
In both E1 and E2 scenarios, performance gradually improves from simple models like Q-RL and PPO toward the more advanced ensemble-based systems. Notably, in \texttt{BSD\_E1\_PA}, the EAAMARL model achieves the best AUC/F1 score of \emph{0.91 / 0.50}, clearly outperforming standalone agents. Across all BSD datasets, EAAMARL consistently outperforms other models with average AUCs exceeding 0.85 and F1-scores above 0.35, showing its robustness to sparse signals typical of BSD traces. AAMARL also performs competitively, especially in PA dataset, indicating the value of active feedback loops.

\paragraph{Windows:}
Detection on Windows datasets exhibits moderate-to-high variance across models. For E1 datasets, \texttt{Windows\_E1\_PA} and \texttt{Windows\_E1\_PX} show top scores by EAAMARL at \emph{0.91 / 0.41} and \emph{0.89 / 0.39} respectively. Interestingly, AE-RL and MARL models are often surpassed by AMARL and AAMARL, emphasizing the benefits of adversarial robustness and active query refinement. In E2 scenarios, performance gains from MARL to EAAMARL are even more pronounced, suggesting stronger generalization under this setup.

\paragraph{Linux:}
Linux datasets reveal that simpler models (Q-RL, PPO) struggle to achieve both reasonable AUC and F1, with most scores hovering around 0.50--0.55. In contrast, MARL and AMARL agents gain momentum, especially in \texttt{Linux\_E1\_PA} and \texttt{Linux\_E2\_PE}, where AAMARL and EAAMARL score up to \emph{0.89 / 0.36}. The autoencoder-based latent representation appears especially valuable in this OS, improving representation learning for RL agents. Overall, EAAMARL demonstrates the most consistent superiority in Linux experiments.

\paragraph{Android:}
Despite Android’s more volatile behavioral traces, ensemble-based agents again lead the pack. \texttt{Android\_E1\_PA} and \texttt{Android\_E2\_PE} highlight this trend, with EAAMARL reaching \emph{0.89 / 0.36} and \emph{0.88 / 0.41}. Q-RL and PPO underperform significantly, with F1 scores often below 0.10, pointing to their poor recall under Android’s dynamic process behaviors. AAMARL and EAAMARL show how adversarial training and active learning can collectively overcome these challenges.
\vspace{-0.8 em}
\paragraph{Overall Trends.}
The general progression in performance across all OSes and scenarios is clear:
\vspace{-0.7 em}
\begin{itemize}
    \item \textbf{QRL, PPO, DQN}: Serve as lower baselines with poor sensitivity and low F1.
    \item \textbf{AE-RL and MARL}: Benefit from better latent representations and multi-agent learning.
    \item \textbf{AMARL and AAMARL}: Introduce robustness and adaptability, especially effective in dynamic or noisy environments.
    \item \textbf{EAAMARL}: Shows state-of-the-art performance in nearly every dataset, combining active learning, adversarial defense, and ensemble synergy.
\end{itemize}

These results support our hypothesis that combining diverse agents with active feedback and latent space modeling significantly enhances APT detection capability across heterogeneous environments.

\vspace{-0.8 em}
\section{Conclusion}
\vspace{-0.8 em}
In this work, we presented a novel architecture for Advanced Persistent Threat (APT) detection that integrates deep autoencoding, multi-agent reinforcement learning (RL), active learning, and ensemble decision-making. We addressed key challenges in the cybersecurity domain—such as the rarity of APT events, concept drift in attacker behavior, and high false-positive costs—by leveraging adaptive and interactive learning agents that evolve through uncertainty-driven feedback. Our proposed framework builds on multiple reinforcement learning paradigms, including Q-Learning, PPO, DQN, adversarial training, and multi-agent collaboration. These agents operate on latent representations extracted from autoencoders, allowing for compact, high-level system behavior abstraction. To further enhance detection sensitivity, we introduced an active learning loop that queries an oracle (simulated from ground truth) whenever uncertainty is detected. This mechanism enables dynamic labeling of ambiguous traces and accelerates agent retraining on the most informative samples. Through extensive evaluation across 40 datasets spanning four operating systems and multiple attack scenarios, we demonstrated that traditional single-agent RL methods perform poorly in isolation. However, performance significantly improves when agents collaborate (Multi-Agent RL), become adversarially trained (AMARL), and incorporate active learning (AAMARL). Our Ensemble Active Adversarial Multi-Agent RL (EAAMARL) framework achieved the best results, consistently outperforming all baselines in terms of AUC and F1 score. The promising performance of EAAMARL confirms the benefit of integrating multiple RL strategies with feedback-aware learning in cybersecurity. As future work, we aim to explore real-time deployment, transfer learning across domains, and the integration of symbolic reasoning or LLMs into the ensemble.
%
%
%
%
\vspace{-0.8 em}
\section*{Disclosure of Interests:} The authors declare that they have no conflict of interest.
 
\bibliographystyle{IEEEtran}
\bibliography{main}
\end{document}